\documentclass[aps, prd, showpacs, superscriptaddress, ctexart, nofootinbib]{revtex4}

\usepackage{graphicx}
\usepackage{latexsym}

\usepackage{color}

\def\be{\begin{equation}}
\def\ee{\end{equation}}
\def\bea{\begin{eqnarray}}
\def\eea{\end{eqnarray}}

\begin{document}

\title{Global fitting analysis on cosmological models after BICEP2}

\author{Hong Li}
\email{hongli@ihep.ac.cn}
\affiliation{Key Laboratory of Particle Astrophysics, Institute of High Energy Physics, Chinese Academy of Science, P.O.Box 918-3, Beijing 100049, P.R.China}

\author{Jun-Qing Xia}
\affiliation{Key Laboratory of Particle Astrophysics, Institute of High Energy Physics, Chinese Academy of Science, P.O.Box 918-3, Beijing 100049, P.R.China}

\author{Xinmin Zhang}
\affiliation{Theoretical Physics Division, Institute of High Energy Physics, Chinese Academy of Science, P.O.Box 918-4, Beijing 100049, P.R.China}


\begin{abstract}
Recently, BICEP2 collaboration has released their results on the
measurements of the CMB polarizations. In the framework of the
$\Lambda$CDM with a power law form of the scalar primordial power
spectrum, this new measurement on the B-mode puts a tight
constraint on tensor/scalar ratio
$r=0.2^{+0.07}_{-0.05}~(1\sigma)$, which however is in tension
with the Planck limit, $r<0.11$ (95\% C.L.). In this paper, we
consider various extensions of the $\Lambda$CDM model by
introducing extra cosmological parameters such as the equation
state of dark energy $w$, the curvature of the universe
$\Omega_k$, the running of the scalar power spectrum index
$\alpha_s$, the sum of the neutrino mass $\Sigma m_{\nu}$, the
effective number of neutrinos $N_{eff}$, the tensor power spectrum
index $n_{t}$, then perform the global fit to the BICEP2, Planck
and also the SN and BAO data. Our results show the cosmological
parameters $\alpha_s$ and $n_t$ are highly and directly correlated
with the tensor/scalar ratio $r$. However indirectly the
parameters $\Omega_k$ and $N_{eff}$ are also correlated with $r$.
We will in this paper give the numerical values on the parameters
introduced and show explicitly how the tension between the BICEP2
and Planck is effectively alleviated by the inclusion of the
parameters $\Omega_k$, $\alpha_s$, $N_{eff}$ and $n_{t}$
separately.

\end{abstract}


\maketitle

\section{Introduction}\label{Int}

Recently, the BICEP2 collaboration announced that they have
detected the primordial tensor perturbation modes at more than
$7\sigma$ confidence level, and it is the most accurate
measurements on the cosmic microwave background (CMB) B-mode
polarization at current stage. Basing on the data, within the
framework of a power-law $\Lambda$CDM, BICEP2 collaboration gives
a tight constraint on the tensor/scalar ratio
as $r=0.2 ^{+0.07}_{-0.05}$\cite{Ade:2014xna}. This is the first
time that data sets a lower bound on the primordial tensor modes
and is the most accurate constraints on tensor/scalar ratio. The
detection of the primordial tensor perturbations plays a crucial
and important role in understanding the early universe and
diagnosing inflation models. The BICEP2 measurements of CMB
polarization power spectrum has a series of interesting
implications on cosmological models \cite{Kehagias:2014wza,
Lizarraga:2014eaa,Brandenberger:2014faa,Cheng:2014bma,
Miranda:2014wga,Gerbino:2014eqa,Wu:2014qxa,Zhang:2014dxk,
xia:2014bounce,Wang:2014kqa}.


The tensor/scalar ratio $r=0.2 ^{+0.07}_{-0.05}$ however is in
tension with the previous constraint from the
Planck\cite{planck_fit}temperature power spectrum,
$r<0.11~(2\sigma)$.
Considering that the tension is given by fitting the data within
the framework of a power-law $\Lambda$CDM model, a more general
data fitting analysis might be possible to provide a way to
alleviate the tension of the two CMB data. In fact, the BICEP2
collaboration in their paper\cite{Ade:2014xna} has pointed out
already that including a running of the scalar power spectrum
index makes both BICEP2 and Planck consistent. The reason for this
is the degeneracies among the cosmological parameters.
In paper\cite{Li:2012ug} we have performed a detailed study on the
impacts of the degeneracies on the determination of the
cosmological parameters. Interestingly we also found the $ 95\% $
upper limit on $r$ changes from $0.15$ in the power-law $\Lambda$
CDM model to $0.37$ when including the running parameter
$\alpha_s$. This result obtained from the fitting with the
previous CMB WMAP7 data together with SN + BAO + HST gives a weak
constraint on $r$, however our analysis shows explicitly how the
upper limit of $r$ gets relaxed when including the running
parameter $\alpha_s$. Last year, when Planck released their data,
their measured value of Hubble constant $H_0$ is shown to be in
tension with
some other measurements given by the lower-redshift methods, such
as the direct $H_0$ probe from the Hubble Space Telescope (HST)
observations of Cepheid variables\cite{hst_riess,hst_freedman}. In
Refs. \cite{Li:2013gka,Xia:2013dea}, by making use of the
parameter degeneracies we have shown that introducing the EoS
parameter of dark energy into the data fitting analysis can be
helpful for reconciling this tension on $H_0$.
 Along this line, in this paper we examine the tension on $r$ between
 BICEP2 and Planck in various extensions of
the power law $\Lambda$CDM model by introducing extra cosmological
parameters such as the equation state of dark energy $w$ , the
curvature of the universe $\Omega_k$, the running of the scalar
power spectrum index $\alpha_s$, the sum of the neutrino mass
$\Sigma m_{\nu}$, the effective number of neutrinos $N_{eff}$, the
tensor power spectrum index $n_{t}$, then perform the global fit.
Our results show that the inflationary parameters $n_t$ and
$\alpha_s$ can be helpful for reconciling the tension. By doing
the fitting analysis, we find that $r$ is negatively correlated
with $\alpha_s$, and the correlation coefficient is around
$-0.35$. The correlation between $r$ and $n_t$ is highly scale
dependent, for pivot scale $k_{pt}=0.05$${\rm Mpc}^{-1}$, they are
positively correlated and the coefficient is $0.78$. So for models
$\Lambda$CDM +$r$+$n_t$ and $\Lambda$CDM +$r$+$\alpha_s$, our
results show the tension can be remarkably reconciled. Our results
also show that even though $N_{eff}$ and $\Omega_k$ are weakly
correlated with $r$, since they are correlated with $n_s$, they
can also soften the tension in certain level.

The structure of our paper is organized as follows. In section II
we will introduce the data sets we adopted in the fitting analysis
and the global fitting procedure. In section III we present the
constraints on various model parameters, and the discussion and
summary are given in section IV.

\section{data sets employed and global fitting analysis}

\subsection{Numerical Method}

We perform the global fitting of cosmological parameters using the
CosmoMC package \cite{cosmomc}, a Markov Chain Monte Carlo (MCMC)
code. The basic cosmological parameters are allowed to vary with
top-hat priors: the cold dark matter energy density parameter
$\Omega_c h^2 \in [0.01, 0.99]$, the baryon energy density
parameter $\Omega_b h^2 \in [0.005, 0.1]$, the curvature term
$\Omega_k \in [-0.3 0.3]$, the effective numbers of neutrinos
$N_{eff} \in [0, 10]$, the total neutrino mass $\Sigma m_{\nu} \in
[0, 1.5]~ eV$, the scalar spectral index $n_s \in [0.5, 1.5]$,
tensor index $n_t \in [-4,5]$, tensor to scalar ratio $r \in [0,
1]$, the primordial amplitude $\ln[10^{10}A_s] \in [2.7, 4.0]$,
the running of primordial scalar power spectrum index $\alpha_s
\in [-0.5, 0.5]$, the ratio (multiplied by 100) of the sound
horizon at decoupling over the angular diameter distance to the
last scattering surface $100\Theta_s \in [0.5, 10]$, and the
optical depth to reionization $\tau \in [0.01, 0.8]$. The pivot
scale\footnote{For discussion on the scale dependence of $n_t$ we
choose another pivot scale of $k_{s0} = 0.001$ ${\rm Mpc}^{-1}$ .}
is set at $k_{s0} = 0.05$ ${\rm Mpc}^{-1}$. In our calculation, we
assume an purely adiabatic initial condition.

The procedure of our study is the following. We start with the
power law $\Lambda$CDM $ + r $ model described by the basic
parameters
\begin{equation}
\left\{ \Omega_bh^2, \Omega_ch^2, \Theta_s, \tau,  n_s,
A_s, r\right\}~.
\end{equation}
Then we consider various extensions of the model above by
introducing the extra cosmological parameters: the equation state
of dark energy $w$, the curvature of the universe $\Omega_k$, the
running of the scalar power spectrum index $\alpha_s$, the sum of
the neutrino mass $\Sigma m_{\nu}$, the effective number of
neutrinos $N_{eff}$ and the tensor power spectrum index $n_{t}$.
And each time, we add only one new parameter for the fitting.

\subsection{Current Observational Data}

In our analysis, we consider the following cosmological probes: i)
power spectra of CMB temperature and polarization anisotropies
released by Planck and polarization E and B modes given by BICEP2
collaborations; ii) the baryon acoustic oscillation in the galaxy
power spectra; iii) luminosity distances of type Ia supernovae.

For the Planck data from the 1-year data release
\cite{planck_fit}, we use the low-$\ell$ and high-$\ell$ CMB
temperature power spectrum data from Planck with the low-$\ell$
WMAP9 \cite{wmap9} polarization data (Planck+WP). We marginalize
over the nuisance parameters that model the unresolved foregrounds
with wide priors \cite{planck_likelihood}, and do not include the
CMB lensing data from Planck \cite{planck_lens}. For the BICEP2
data sets, we adopt 9 bins of the E and B polarization field data
sets in range of $\ell [30,150]$\footnote{In the current study, we
focus on the constraint on the primordial tensor perturbation, so
have not included the POLARBEAR\cite{Ade:2014afa} results.}.

Baryon Acoustic Oscillations provides an efficient method for
measuring the expansion history by using features in the
clustering of galaxies within large scale surveys as a ruler with
which to measure the distance-redshift relation\cite{bao}. It measures
not only the angular diameter distance, $D_A(z)$, but also the
expansion rate of the universe, $H(z)$, which is powerful for
studying dark energy \cite{task}. Since the current BAO data are
not accurate enough for extracting the information of $D_A(z)$ and
$H(z)$ separately \cite{okumura}, one can only determine an
effective distance \cite{baosdss}:
\begin{equation}
D_V(z)=[(1+z)^2D_A^2(z)cz/H(z)]^{1/3}~.
\end{equation}
Following the Planck analysis \cite{planck_fit}, in this paper we
use the BAO measurement from the 6dF Galaxy Redshift Survey
(6dFGRS) at a low redshift ($r_s/D_V (z = 0.106) = 0.336\pm0.015$)
\cite{6dfgrs}, and the measurement of the BAO scale based on a
re-analysis of the Luminous Red Galaxies (LRG) sample from Sloan
Digital Sky Survey (SDSS) Data Release 7 at the median redshift
($r_s/D_V (z = 0.35) = 0.1126\pm0.0022$) \cite{sdssdr7}, and the
BAO signal from BOSS CMASS DR9 data at ($r_s/D_V (z = 0.57) =
0.0732\pm0.0012$) \cite{sdssdr9}.

Finally, we include data from Type Ia supernovae, which consists
of luminosity distance measurements as a function of redshift,
$D_L(z)$. In this paper we use the supernovae data set,
``Union2.1'' compilation, which includes 580 supernovae
reprocessed by Ref.\cite{union2.1}. When calculating the
likelihood, we marginalize the nuisance parameters, like the
absolute magnitude $M$.

\section{Constraints on cosmological models }
\begin{figure}
\begin{center}
\includegraphics[scale=0.33]{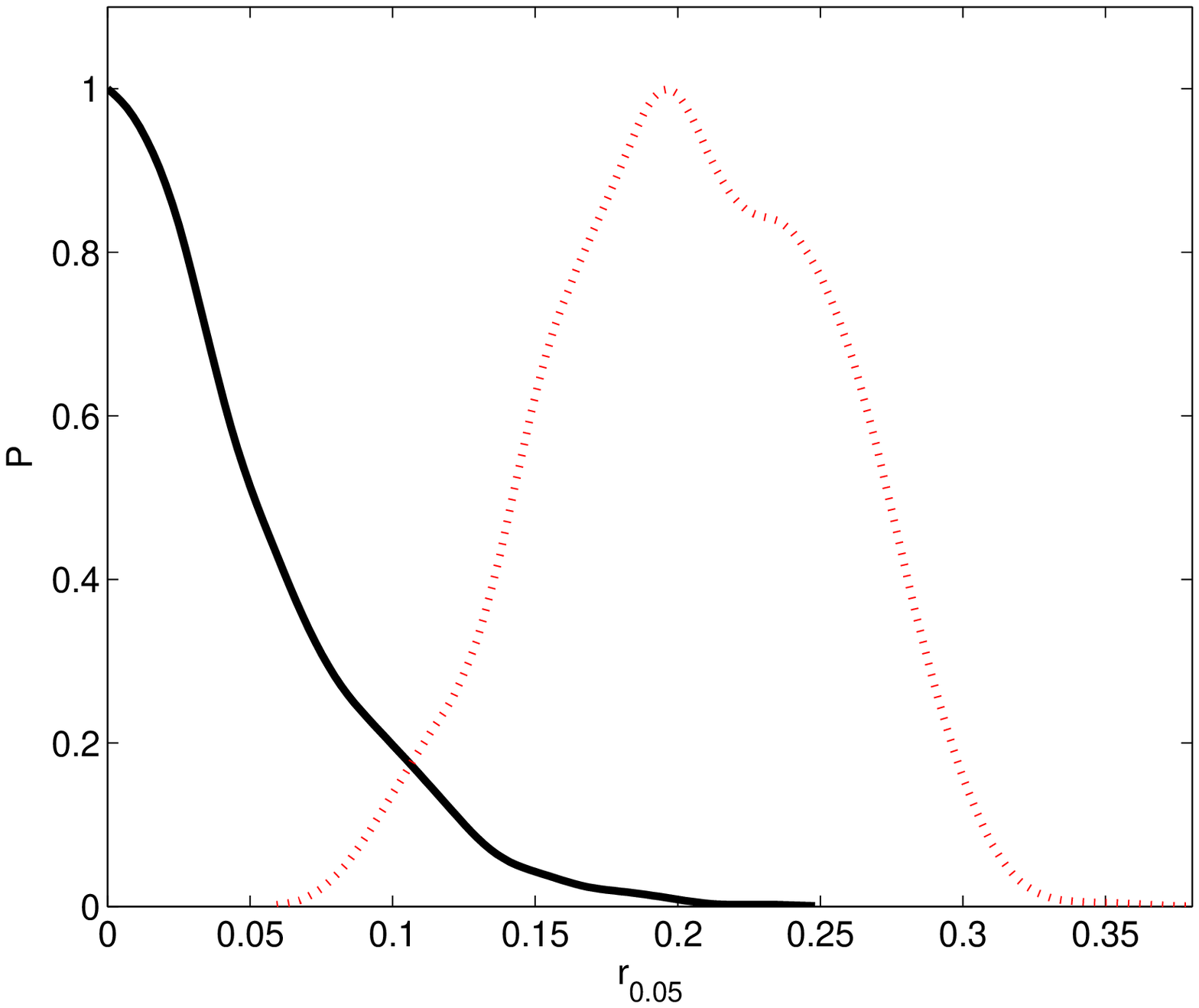}
\includegraphics[scale=0.50]{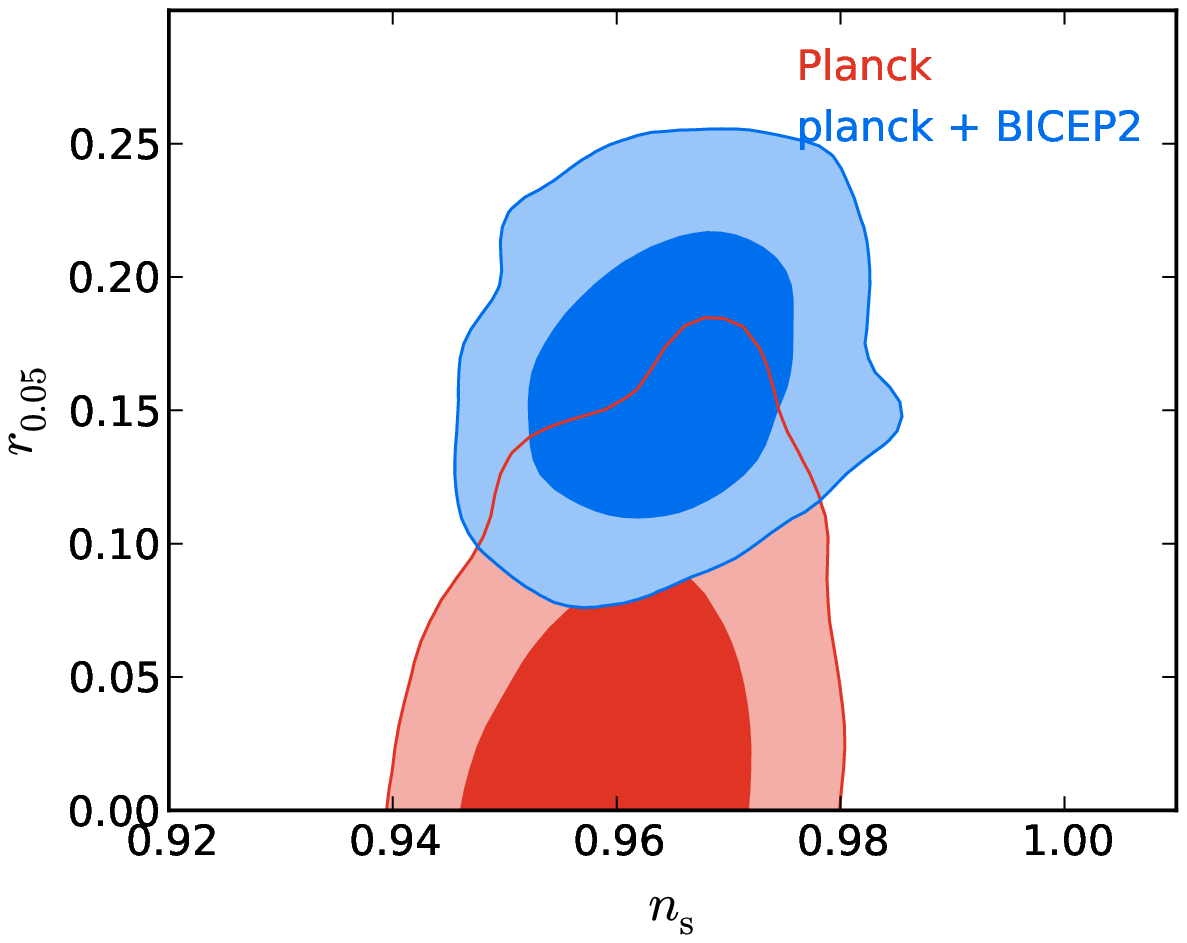}
\caption{The left panel: One-dimensional marginalized probability
distribution of $r$. The red line is given by BICEP2 and the black
line is given by Planck. The right panel: two-dimensional cross
correlation between $r$ and $n_s$. }\label{1d2d-r}
\end{center}
\end{figure}

We start to present our numerical results with only the CMB data.
Within the power law $\Lambda$CDM model, we show in the left panel
of figure \ref{1d2d-r} the one dimensional probability
distribution of $r$ given by fitting with planck temperature and
BICEP2 polarization E and B modes respectively. From Planck it
gives $r< 0.11~ (1\sigma)$ and BICEP2 give $r=
0.2^{+0.07}_{-0.05}~ (1\sigma)$. We also show the comparison of
the constraints in two dimensional plane of $r$-$n_s$ in the right
panel of figure \ref{1d2d-r} and one can see that there's no
overlap for the two $68\%$ confidence level contours.


\begin{table}{\footnotesize
\caption{ We list the constraints on $r$ and the new parameters
introduced in extensions of the power law $\Lambda$CDM model from
Planck+BICEP2 and also Planck only for comparison. The minimal
$\chi^2$ listed are for the corresponding models when fitting with
Planck+BICEP2. The correlation coefficients indicated by ($corr$)
show the correlation between $r$ and the new parameter introduced
 in fitting with data sets of Planck+SN+BAO+BICEP2. }\label{t2}
\begin{center}

\begin{tabular}{c|c|c|c|c|c|c|c}

\hline\hline
&$\Lambda$CDM+$r$&$\Lambda$CDM+$r$+$n_t$&$\Lambda$CDM+$r$+$\alpha_s$&$\Lambda$CDM+$r$+$N_{eff}$&$\Lambda$CDM+$r$+$\Omega_k$&$\Lambda$CDM+$r$+$\Sigma m_{\nu}$&$\Lambda$CDM+$r$+$w$ \\
\hline
PLK+BCP2&$r=0.165\pm0.036$&$n_t=0.79\pm0.21$&$\alpha_s=-0.027\pm0.01$&$N_{eff}=4.03^{+0.43}_{-0.48}$&$\Omega_k=-0.055^{+0.051}_{-0.024}$&$\Sigma m_{\nu}<0.616~eV$&$w=-1.55^{+0.18}_{-0.34}$\\
\hline
PLK only&$r<0.117$&$no limit$&$\alpha_s=-0.221^{+0.011}_{-0.0009}$&$N_{eff}=3.42\pm0.6$&$\Omega_k=-0.051^{+0.035}_{-0.028}$&$\Sigma m_{\nu}<0.813~eV$&$w=-1.509^{+0.226}_{-0.415}$\\
\hline
$\chi^2_{min}$&$9853.3$&$9845$&$9846.7$&$9848.2$&$9847.0$&$9853.7$&$9853.6$\\
\hline
$corr$&$-$&$0.78$&$-0.35$&$0.05$&$-0.18$&$0.07$&$-0.03$\\
\hline\hline
\end{tabular}
\end{center}}
\end{table}

Now we consider various extensions of the power law $\Lambda$CDM
model. In table \ref{t2}, we list the constraints on the relevant
parameters from Planck + BICEP2 and also the values from Planck
only. For example, the constraints on the EoS of dark energy from
BICEP2 + Planck is $w=-1.55^{+0.18}_{-0.34}$, and with Planck
only, $w=-1.509^{+0.226}_{-0.415}$. This shows with BICEP2 +
Planck, the errors get smaller and the central value of $w$
becomes more negative. For the curvature of the universe,
$1\sigma$ limit is $\Omega_k=-0.055^{+0.051}_{-0.024}$, the total
neutrino mass $\Sigma m_{\nu}<0.616~eV$, the effective number of
neutrinos $N_{eff}=4.03^{+0.43}_{-0.48}$, all of which are
consistent and slightly tighten the constraints from Planck.

\begin{table}
\caption{ The constraints on tensor/scalar ratio $r$ and the relevant extra cosmological parameters from Planck+SN+BAO with/without BICEP2. For Planck+SN+BAO we present
the 95\% C.L. upper limit, and for Planck+SN+BAO+BICEP2 we list
the $1\sigma$ constraints.}\label{t1}
\begin{center}

\begin{tabular}{c|c|c|c|c}

\hline\hline
models&\multicolumn{2}{|c|}{$Planck+SN+BAO$}&\multicolumn{2}{|c}{$Planck+SN+BAO+BICEP2$}\\
\hline
$\Lambda$ CDM+$r$&$r~<~0.12$&$-$&$0.165^{+0.031}_{-0.040}$&$-$\\
$\Lambda$ CDM+$r$+$n_t$&no-limit&$n_t>0.296$&$r>0.370$&$n_t=0.822^{+0.240}_{-0.182}$\\
$\Lambda$ CDM+$r$+$\alpha_s$&$r~<~0.25$&$\alpha_s=-0.022\pm{0.011}$&$r=0.199^{+0.037}_{-0.044}$&$\alpha_s=-0.0281\pm0.0099$\\
$\Lambda$ CDM+$r$+$N_{eff}$&$r~<~0.15$&$N_{eff}=3.503^{+0.275}_{-0.274}$&$r=0.169^{+0.031}_{-0.038}$&$N_{eff}=3.678^{+0.299}_{-0.323}$\\
$\Lambda$ CDM+$r$+$\Omega_k$&$r~<~0.14$&$100\Omega_k=-0.0731\pm0.34$&$r=0.173^{+0.035}_{-0.041}$&$100\Omega_k=-0.25\pm0.34$\\
\hline\hline
\end{tabular}
\end{center}
\end{table}

\begin{figure}
\begin{center}
\includegraphics[scale=0.50]{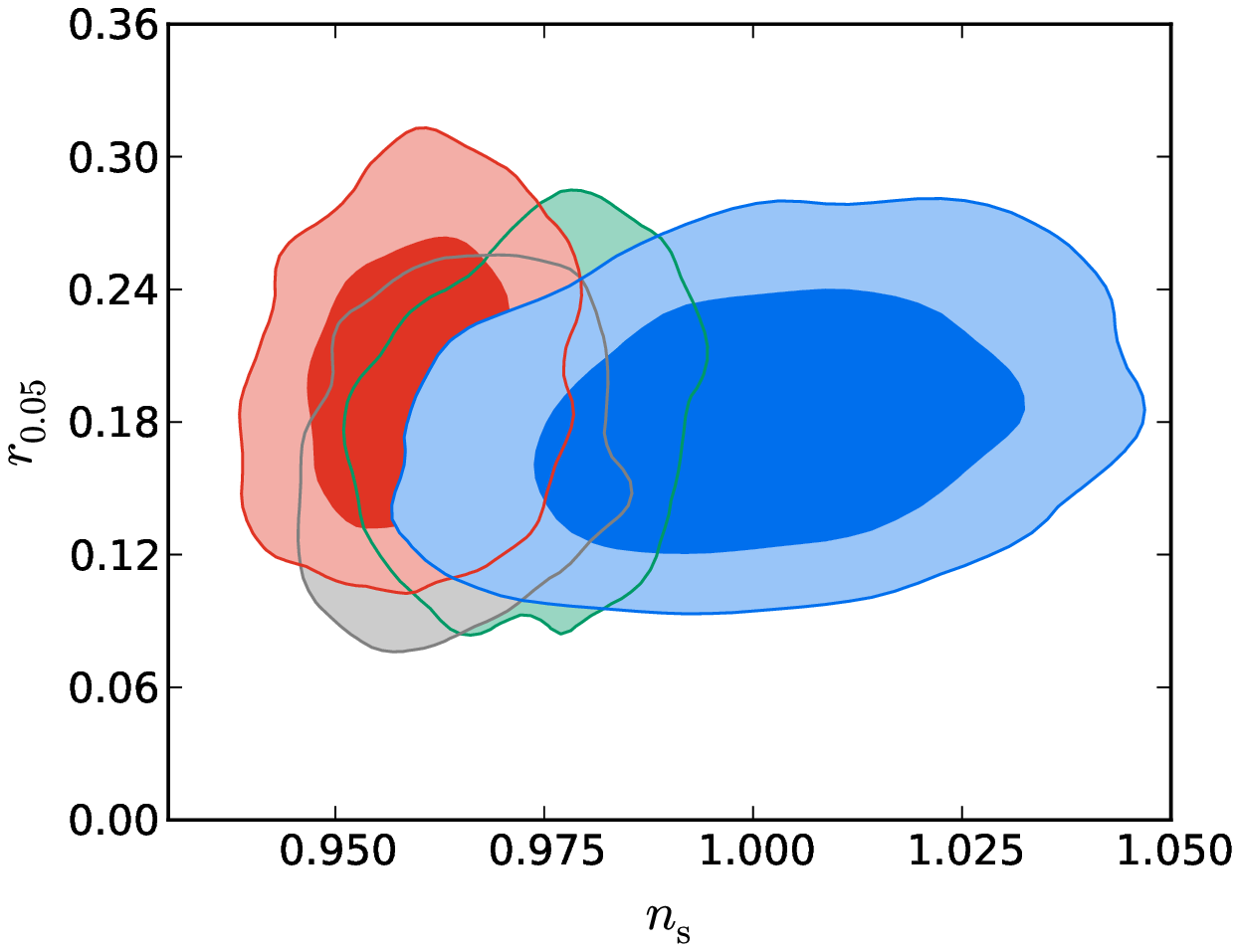}
\includegraphics[scale=0.50]{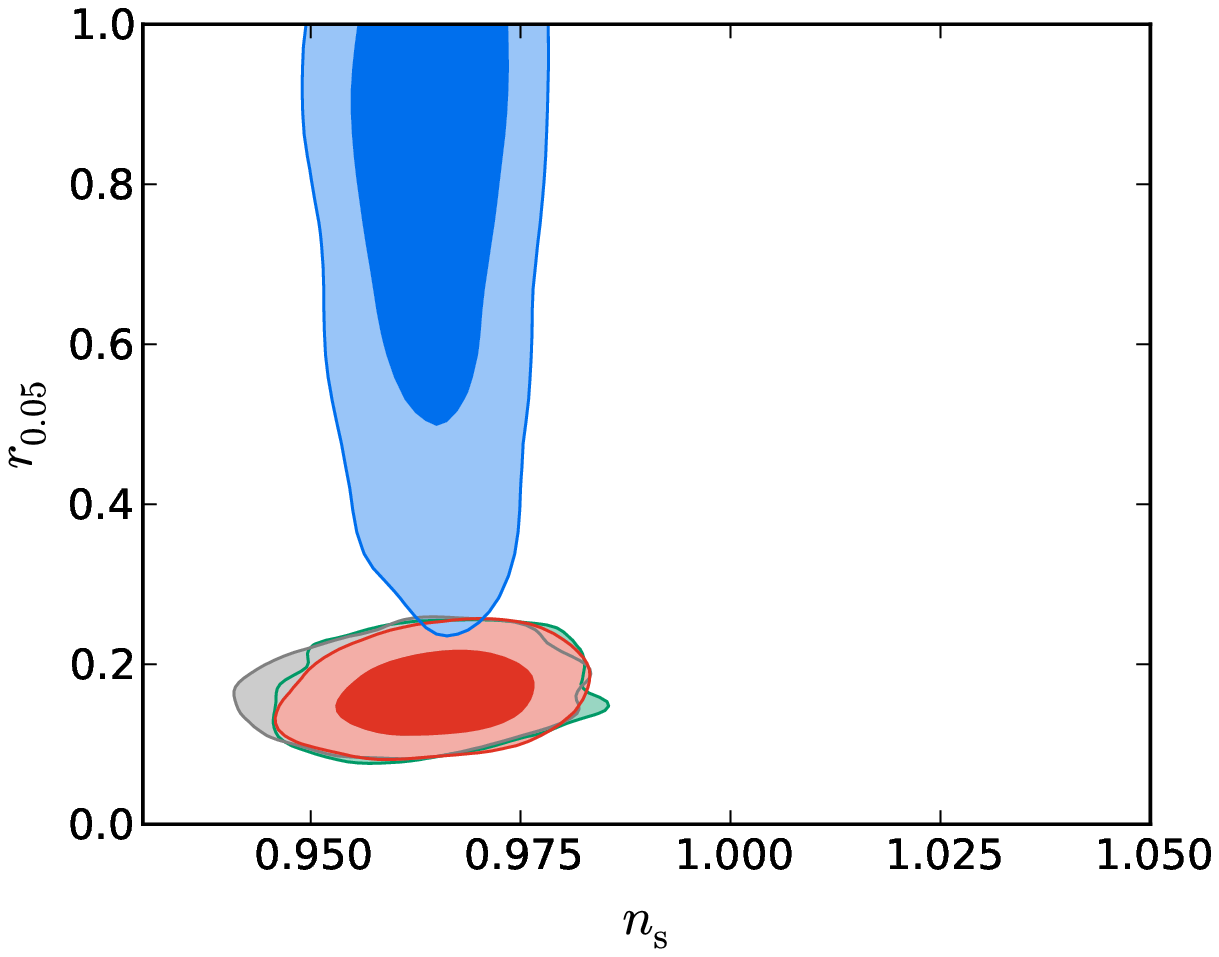}
\caption{The two-dimensional cross correlation between $r$ and $n_s$ by fitting with Planck $+$ BICEP2. In the left panel we compare models of $\Lambda$CDM$+r$ (gray), $\Lambda$CDM$+r+\alpha_s$ (red), $\Lambda$CDM$+r+N_{eff}$ (blue) and $\Lambda$CDM$+r+\Omega_k$ (green), respectively. In the right panel we compare models of $\Lambda$CDM$+r$ (green), $\Lambda$CDM$+r+\Sigma m_{\nu}$ (gray), $\Lambda$CDM$+r+w$ (red), and $\Lambda$CDM$+r+n_t$ (blue), respectively.
}\label{r-ns-c5}
\end{center}
\end{figure}

In figure \ref{r-ns-c5}, we show the two dimensional cross
correlation between $r$ and $n_s$ in different models by fitting
with Planck + BICEP2. We compare the models of $\Lambda$CDM
$+r+\alpha_s$, $\Lambda$CDM $+r+N_{eff}$, and $\Lambda$CDM
$+r+\Omega_k$ in the left panel, while in the right panel we
compare $\Lambda$CDM $+r+\Sigma m_{\nu}$, $\Lambda$CDM $+r+w$ and
$\Lambda$CDM $+r+n_t$. One can see clearly the effects of
$\alpha_s$, $n_t$, $N_{eff}$ and $\Omega_k$. However, for the
parameters $\Sigma m_{\nu}$ and $w$, the effects they can bring
are minor. Thus in the study below, we will focus on the 4 types
of models which corresponds to the inclusion of extra parameters
of $\alpha_s$, $n_t$, $N_{eff}$ and $\Omega_k$, respectively.

\begin{figure}
\begin{center}
\includegraphics[scale=0.50]{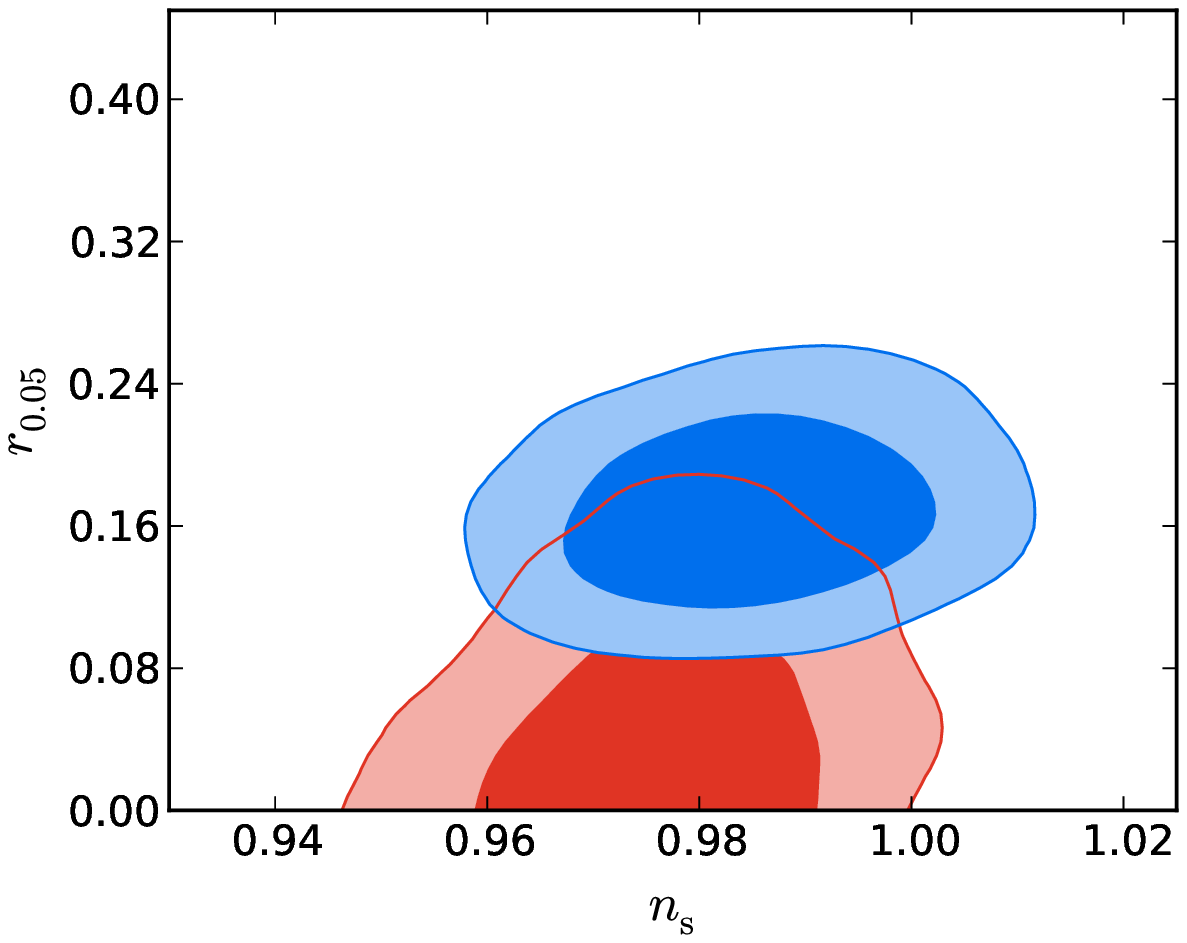}
\includegraphics[scale=0.50]{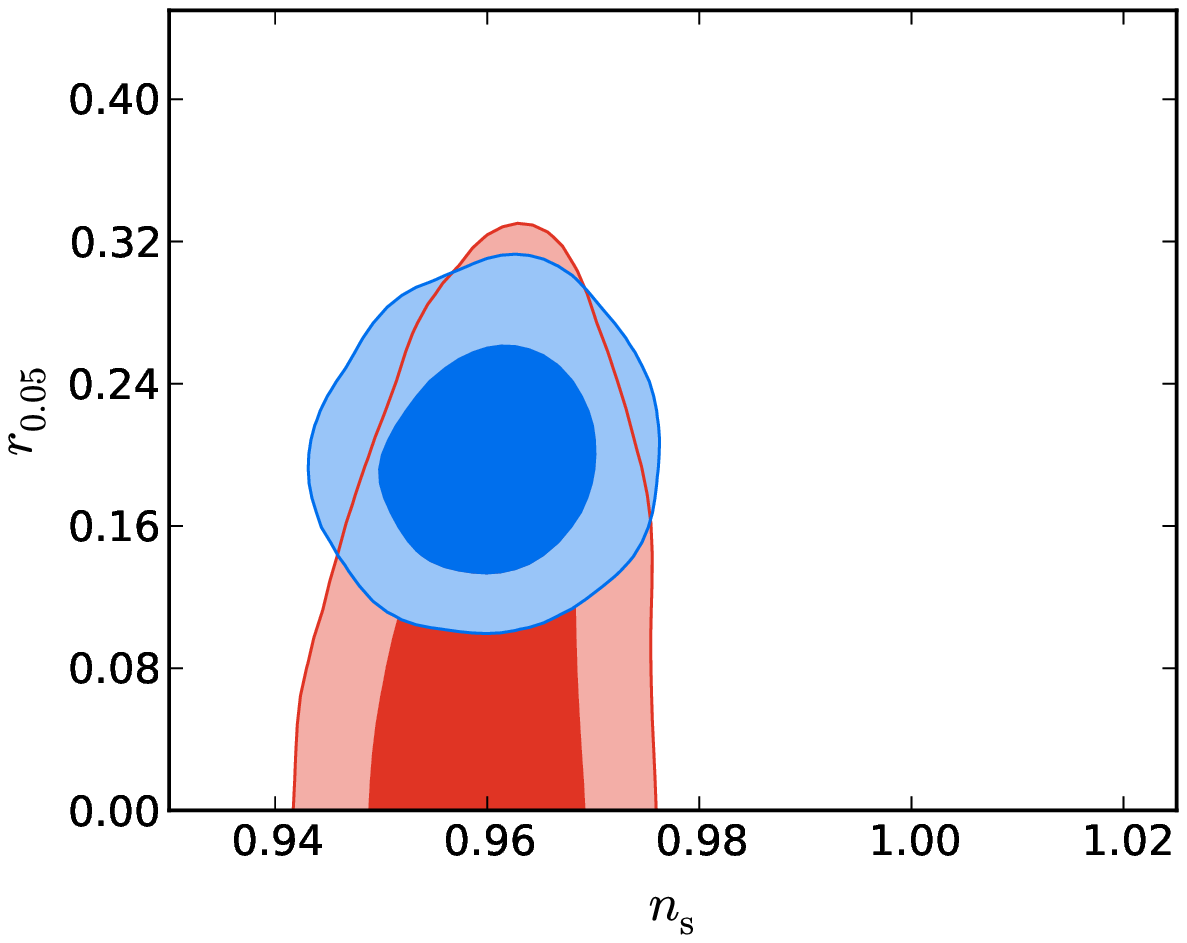}
\includegraphics[scale=0.50]{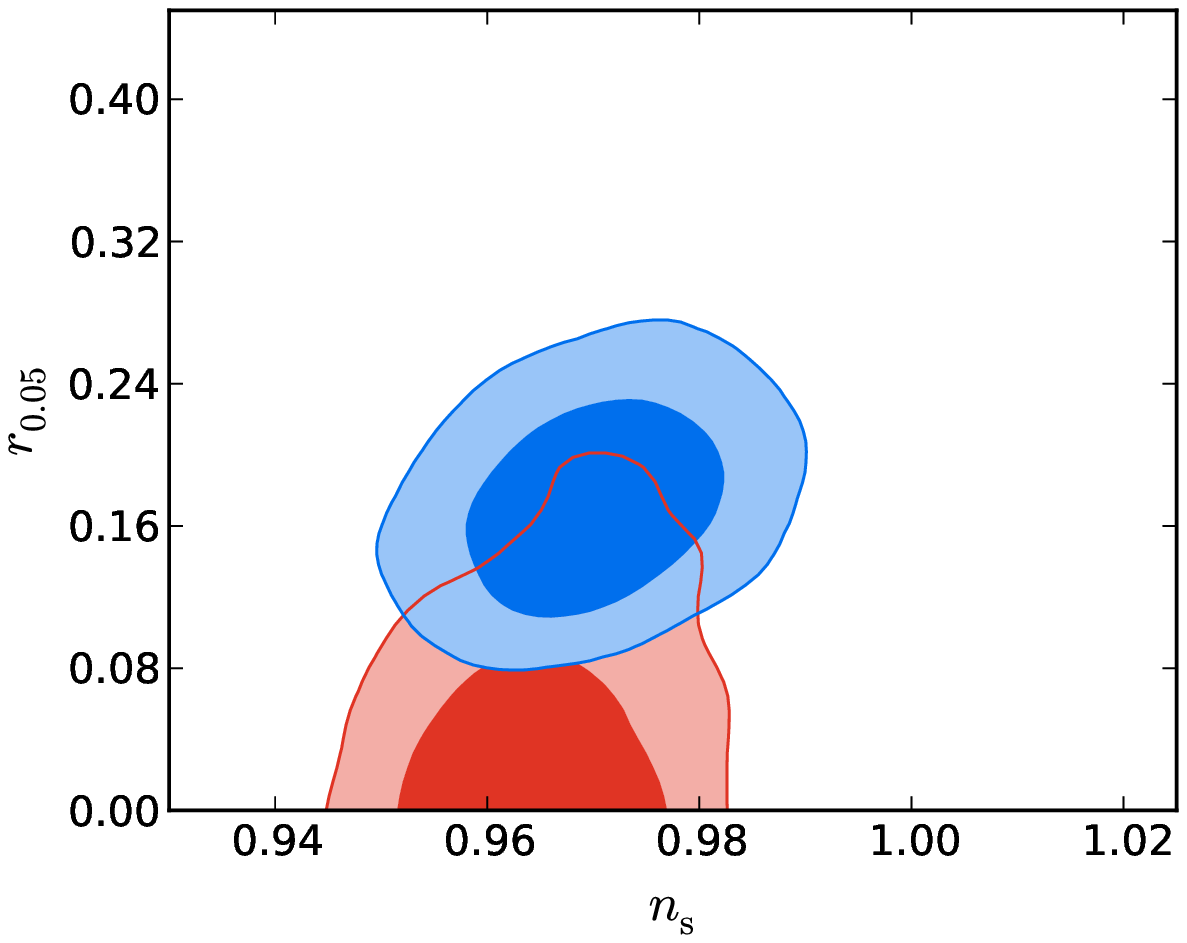}
\includegraphics[scale=0.50]{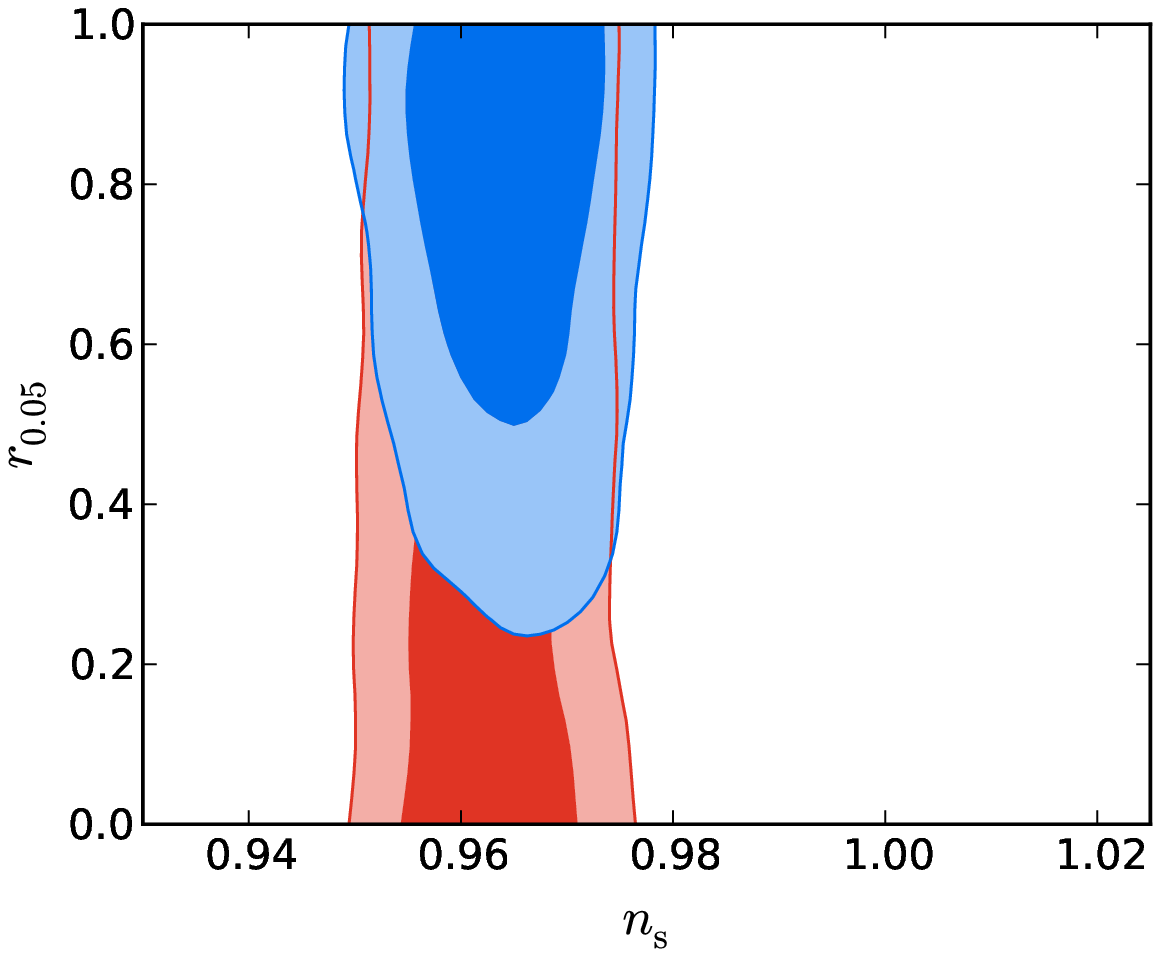}
\caption{Two-dimensional cross correlation between $r$ and $n_s$
by fitting with data sets of Planck + SN + BAO  with (blue)
and without (red) BICEP2. The panels are given by
involving parameters $N_{eff}$ (upper left), $\alpha_s$ (upper right), $\Omega_k$ (lower left) and $n_t$ (lower right)
respectively. }\label{r-ns-4m}
\end{center}
\end{figure}

In order to get a tight constraint on the parameters, we now
combine the CMB data with SN and BAO. We compare the $1$ and
$2\sigma$ confidence contours in plane of $r$-$n_s$ for the 4
different cosmological models in figure \ref{r-ns-4m} and table
\ref{t1}. The contours in red are given by Planck + SN + BAO,
while blue is given by Planck + SN + BAO + BICEP2.

Firstly, we study the effect from $n_t$. During the fitting
procedure we free $n_t$ and $r$ simultaneously and ignore the
inflation consistency relation $n_t= -\frac{A_t}{8A_s}$. We find
that for the pivot scale $k_{pvot}=0.05$ the tensor index is
positively correlated with $r$, and the correlation coefficient
between them is about $0.78$. Such a high correlation leads to a
very weak constraint on both $r$ and $n_t$, which one can see from
Planck + SN + BAO. By taking into account the BICEP2 data, there's a
lower limit on $r$, and $r>0.370$, the constraint on $n_t$ is
$n_t=0.822^{+0.240}_{-0.182}$. The constraints are pivot scale
dependent. In figure \ref{dfpvt} we plot the results with two
pivot scales $k_{pt}=0.001$ $Mpc^{-1}$ and $k_{pt}=0.004$
$Mpc^{-1}$ for comparison. One can see for $k_{pt}=0.001$
$Mpc^{-1}$, we get constraints like $r_{0.001}<0.076$ and
$n_{t_{0.001}}=1.04\pm0.39$, and for $k_{pt}=0.004$ $Mpc^{-1}$,
the constraints give $r_{0.004}=0.094\pm0.058$ and
$n_{t_{0.004}}=0.94\pm0.76$.

\begin{figure}
\begin{center}
\includegraphics[scale=0.40]{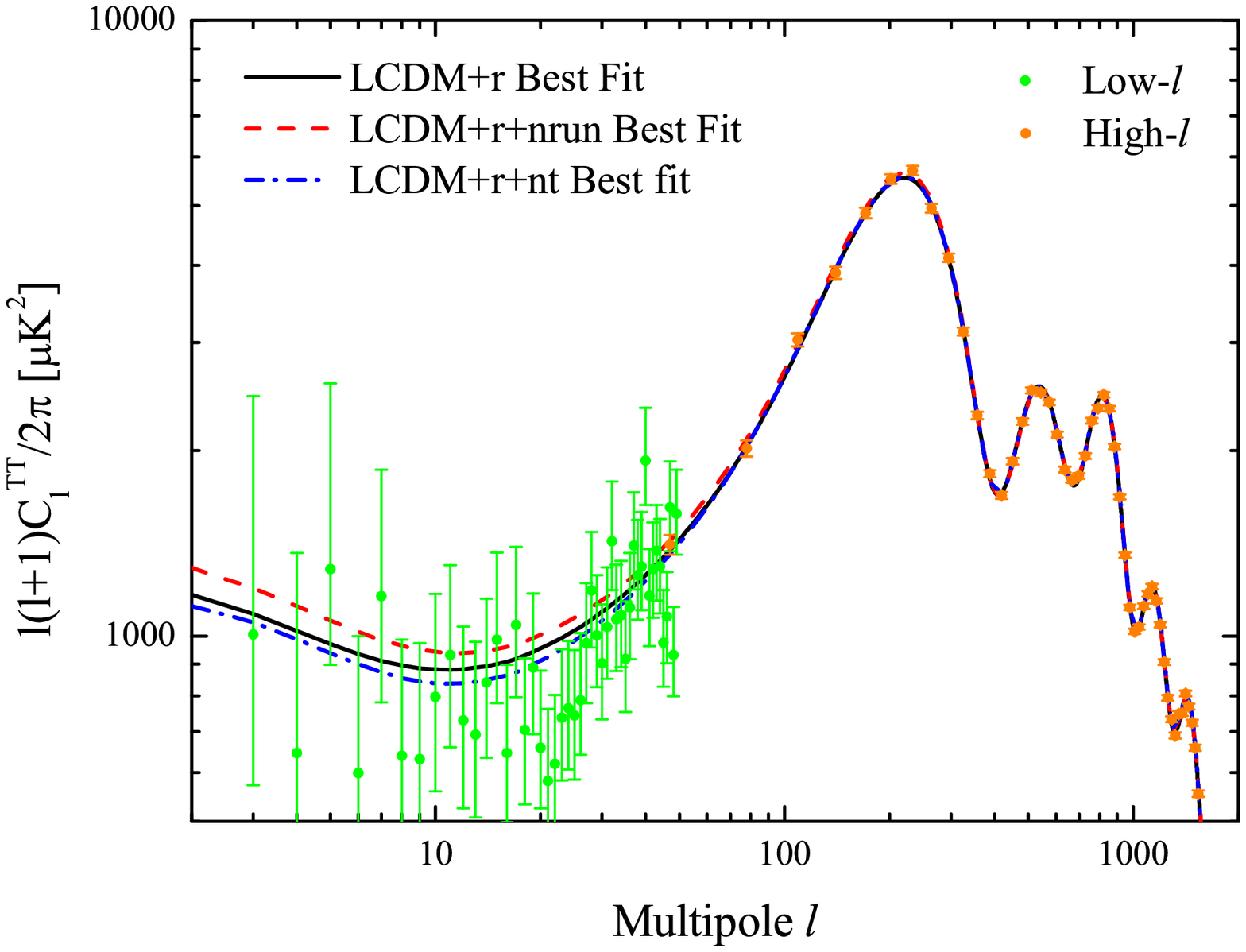}
\includegraphics[scale=0.40]{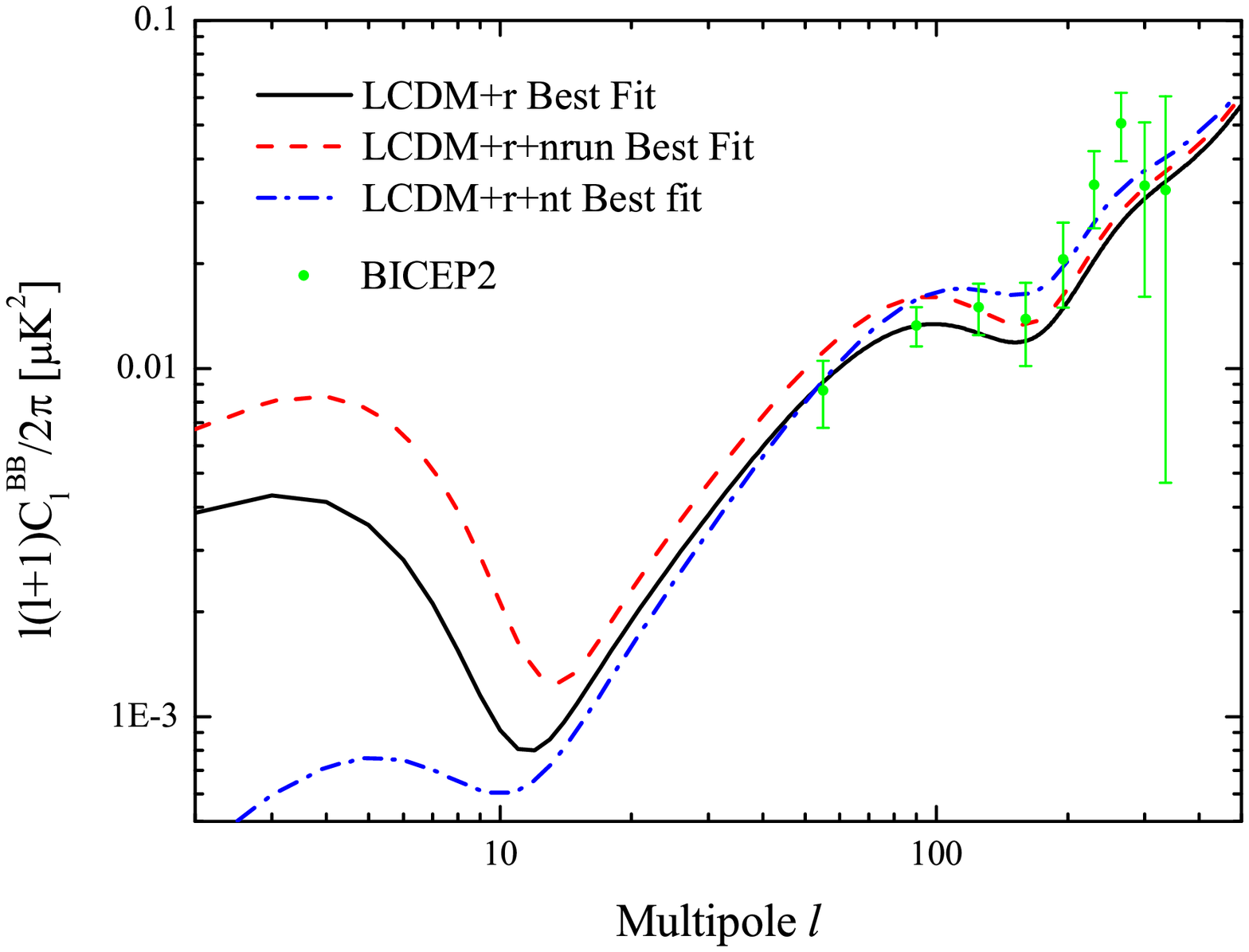}
\caption{Theoretical CMB  TT  ({\it Left}) and BB ({\it Right}) power spectra for the best fit $\Lambda$CDM$+r$, $\Lambda$CDM$+r+n_t$ and  $\Lambda$CDM$+r+\alpha_s$ models, as well as the Planck and BICEP2 observational data.
}\label{comp-TTBB}
\end{center}
\end{figure}

\begin{figure}
\begin{center}
\includegraphics[scale=0.50]{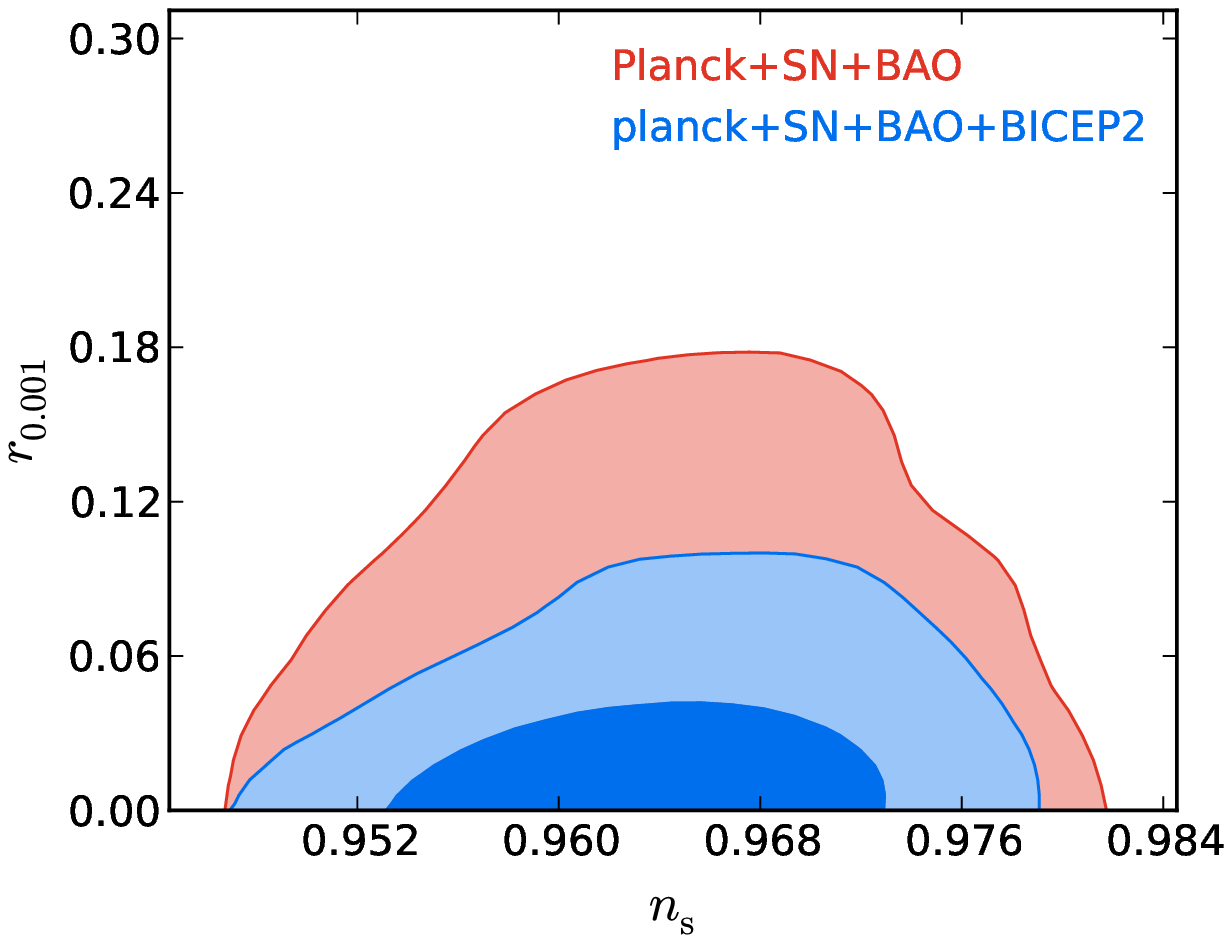}
\includegraphics[scale=0.50]{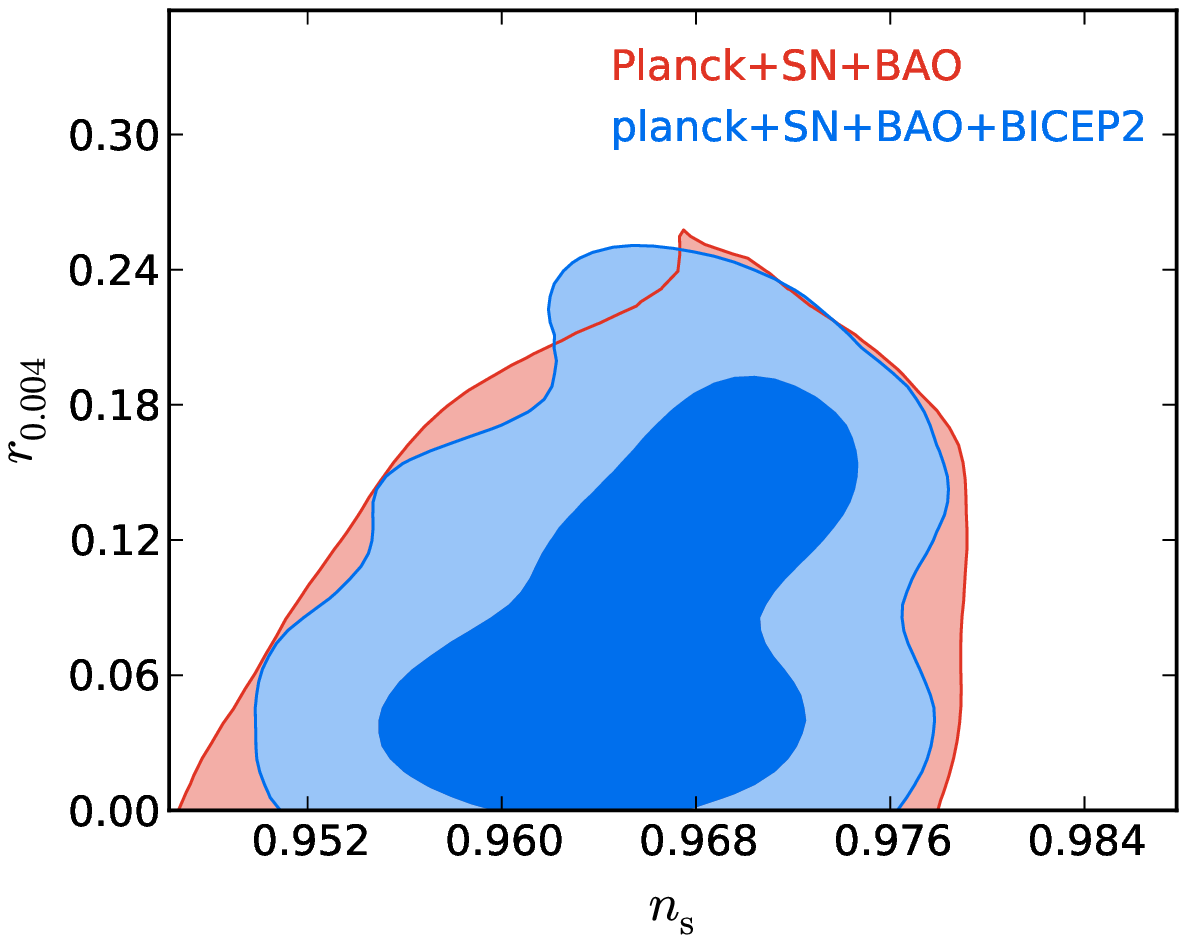}
\caption{Two-dimensional cross correlation constraints of $r$ and $n_s$ at pivot scales $k_{pt}=0.001$${\rm Mpc}^{-1}$ and $k_{pt}=0.004$${\rm Mpc}^{-1}$ by fitting with data sets of Planck + SN + BAO with (blue) and without (red) BICEP2.
}\label{dfpvt}
\end{center}
\end{figure}

Then we consider the correlations between $r$ and the running
parameter $\alpha_s$. Our results are shown in the upper right
panel of figure \ref{r-ns-4m}. The effect from the running is also
sizable, as shown in the figure, since $\alpha_s$ is negatively
correlated with $r$. For a negative running which is preferred by
the current data, a larger $r$ can satisfy both Planck and BICEP2.
In this case, we get $r=0.199^{+0.037}_{-0.044}$ with data sets of
Planck+SN+BAO+BICEP2, and $\alpha_s=-0.0281\pm0.0099$ which give a
negative running of $n_s$. By comparing with $\Lambda$CDM  $+r$,
including the running parameter can decrease the minimal $\chi^2$
of the fitting about $7$, which indicates that $\alpha_s$ can be
helpful for balancing the two CMB data. In figure \ref{comp-TTBB},
we plot the expected TT and BB power spectra from the best fit
models of $\Lambda$CDM$+r$, $\Lambda$CDM$+r+n_t$ as well as
$\Lambda$CDM$+r+\alpha_s$. One can see from the BB power spectrum,
the different models can account for the BICEP2 data equally well,
but behave differently at small $\ell$, which can be distinguished
by the future CMB measurements.

The effects from $N_{eff}$ and $\Omega_k$ are shown in the upper
left panel and lower left panel of figure \ref{r-ns-4m}
respectively. We find that $N_{eff}$ and $\Omega_k$ are very
weakly correlated with $r$ and the coefficients are both around
$0$. However, they are strongly correlated with $n_s$, so they can
indirectly affect the allowed parameter space. Our results show
these two parameters can also be helpful to compromising the
tension in certain level. Numerically we get
$r=0.169^{+0.031}_{-0.038}$ and $r=0.173^{+0.035}_{-0.041}$ by
introducing $N_{eff}$ and $\Omega_k$ respectively. And the
constraints on $N_{eff}$ and $\Omega_k$ are
$N_{eff}=3.678^{+0.299}_{-0.323}$ and $100\Omega_k=-0.25\pm0.34$,
which are consistent with those from Planck and other
observations.


Finally, we use the new BICEP2 data to study the rotation angle
induced by the CPT-violating interaction \cite{feng:2006}. Here,
we only consider the isotropic rotation $\alpha$ and leave the
analysis on the direction-dependent rotation angle
$\alpha(\hat{\bf n})$ in the future work.

As we know, due to the non-zero rotation angle, the CMB
polarization power spectra are modified. Especially, we have the
non-zero CMB TB and EB cross power spectra, which vanish in the
standard CMB theory. Therefore, we have used the published CMB TB
and EB observational data, such as the WMAP and BICEP experiments,
to constrain the rotation angle and test the CPT symmetry.

In 2010, the BICEP experiment published the two-year BICEP data,
including the TB and EB information \cite{bicep1:2year}. In ref.
\cite{xia:2010cpt} we analysed this data and found that a non-zero
rotation angle is favored at about $2.4\sigma$ confidence level,
$\alpha=-2.60\pm1.02$ deg (68\% C.L.), due to the clear bump
structure in the BICEP TB and EB power spectra at $\ell \sim 150$.
In 2013, the BICEP collaboration released the three-year data and
re-analysed the constraint of the rotation angle
\cite{bicep1:3year}. They show when using the standard calibration
method, the BICEP data still prefer a non-vanishing TB and EB
power spectra with an overall polarization rotation $\alpha=-2.77
\pm 0.86 ({\rm stat.}) \pm 1.3 ({\rm sys.})$ deg (68\% C.L.),
which still implies $\alpha\neq0$ at $\sim 2\sigma$ confidence
level \cite{bicep1:3year,li:2014cpt}.

\begin{figure}
\begin{center}
\includegraphics[scale=0.45]{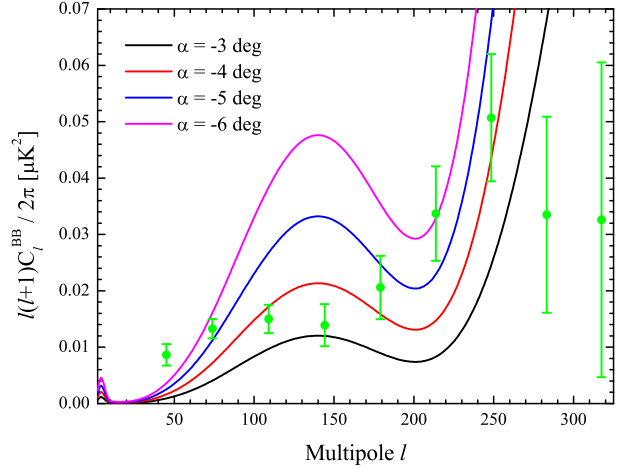}
\caption{Theoretical predictions of the CMB BB power spectra from the non-zero rotation angles and comparison with the BICEP2 BB data.
}\label{bicep:bb}
\end{center}
\end{figure}

A sizable rotation angle will have an interesting effect on the CMB
BB power spectrum (see Fig.4 of our paper in \cite{xia:2010cpt}).
Here, in figure \ref{bicep:bb} we plot the CMB BB power spectra
with several rotation angles. The BICEP2 BB data are also shown
there. One can see a rotation angle $\sim 4$ deg will generate a
B mode with an amplitude at the order of magnitude comparable
to the BICEP2. Some recent studies on the implications of rotation
angle\cite{rta_Li2008tma} on the BICEP2 results can be found in
\cite{zhao&li,liu:cpt}.

In ref. \cite{bicep1:3year} the authors proposed a new calibration
method, ``self-calibration", in which they used the derived
rotation angle to calibrate the detector polarization
orientations. With the self calibration method, the rotation angle
is minimized and the constraint on the tensor-to-scalar ratio $r$
as shown in \cite{bicep1:3year} becomes tighten from $r<0.70$ to
$r<0.65$ at 95\% confidence level. With the new BICEP2 data on EB
and TB, we have analysed the rotation angle and obtained
numerically  $\alpha=0.12\pm 0.16$ deg at the 68\% C.L. which is
expected and consistent with the use of "self calibration".
However it is curious to estimate how large the rotation angle
could be before the "self calibration". To do this, we recall
BICEP2 collaboration in their paper ( see section 8.2) states they
have done a rotation of $\sim 1$ deg, based on which we obtain a
conservative value of the rotation angle before self calibration
which is $ \sim 0.88$ deg . Assuming the error $0.16$ deg remains
almostly the same before and after the self calibration, we
estimate the rotation angle before the self calibration is around
$\alpha= 0.88 \pm 0.16$ deg which is non zero at more than
5$\sigma$ significance.

\section{Summary and discussion}\label{Sum}

Recently, the BICEP2 collaboration published their result on
primordial tensor modes from the highest resolution E and B maps
of the CMB polarization anisotropy. In the standard $\Lambda$CDM
model, the BICEP2 data give a tight constraint on $r$, which
prefers a larger tensor mode when comparing with the Planck
temperature measurements. In this paper we have considered several
extensions of the $\Lambda$CDM models, and found that 4 models by
introducing extra cosmological parameters, $n_t$, $\alpha_s$,
$N_{eff}$ and $\Omega_k$ can alleviate this tension.
 Furthermore, we have performed a detailed numerical calculation
and fitted the model parameters to the BICEP2 + Planck + SN + BAO.
On constraining the tensor index $n_t$, we find that it is pivot
scale dependent, and with BICEP2 data, a blue tilt of the
primordial gravitational waves spectrum can be allowed. Our study
will be important to understand the resolution of the tension on
$r$ between BICEP2 and Planck, and very useful to the model
building in resolving this tension.

\section*{Acknowledgements}

We acknowledge the use of the Legacy Archive for Microwave
Background Data Analysis (LAMBDA). We thank Antony Lewis, Zuhui
Fan, Qing-guo Huang, Meng Su and Gongbo Zhao for helpful
discussion. HL is supported in part by the National Science
Foundation of China under Grant Nos. 11033005 and 11322325, by the
973 program under Grant No. 2010CB83300. J.X. is supported by the
National Youth Thousand Talents Program. X.Z. is supported in part
by the National Science Foundation of China under Grants Nos.
11121092, 11033005 and 11375202. This paper is supported in part
by the CAS pilotB program.


\end{document}